\documentclass[11pt]{article}
\usepackage{amsfonts}
\usepackage{amsfonts}
\usepackage{amsfonts,amsmath,amssymb,epsf}
\usepackage{graphicx,color}
\usepackage[usenames,dvipsnames]{pstricks}
\usepackage{epsfig}
\usepackage{pst-grad} % For gradients
\usepackage{pst-plot} % For axes
\usepackage{verbatim}
\usepackage{cite}

\newcommand{\de}{\partial}
\newcommand{\be}{\begin{equation}}
\newcommand{\ba}{\begin{eqnarray}}
\newcommand{\ea}{\end{eqnarray}}
\newcommand{\ee}{\end{equation}}

\newcommand{\we}{\wedge}

\newcommand{\f}{\frac}
\newcommand{\s}{\sqrt}

\newcommand{\ti}{\tilde}
\newcommand{\ap}{\alpha}

\newcommand{\ddd}{\cdot\cdot\cdot}
\newcommand{\no}{\nonumber \\}

\newcommand{\ep}{\epsilon}

\newcommand{\non}{\nonumber}
\begin{document}

\begin{titlepage}
\thispagestyle{empty}

\begin{flushright}
IPMU09-0090 \\
KUNS-2224
\end{flushright}

\bigskip
\begin{center}
\noindent{\large \textbf{\!\!\!Some No-go Theorems for String Duals
of  Non-relativistic  Lifshitz-like  Theories\!\!\!}}\\
\vspace{15mm} \
 Wei Li$^{a}$\footnote{e-mail:
wei.li@ipmu.jp},
Tatsuma Nishioka$^{b,a}$\footnote{e-mail: nishioka@gauge.scphys.kyoto-u.ac.jp} and
Tadashi Takayanagi$^{a}$\footnote{e-mail: tadashi.takayanagi@ipmu.jp}\\
\vspace{1cm}
{\it $^{a}$Institute for the Physics and Mathematics of the Universe (IPMU), \\
 University of Tokyo, Kashiwa, Chiba 277-8582, Japan\\
$^{b}$Department of Physics, Kyoto University, Kyoto 606-8502, Japan}

\vskip 3em
\end{center}

\begin{abstract}
We study possibilities of string theory embeddings of the gravity
duals for non-relativistic Lifshitz-like theories with anisotropic
scale invariance. We search classical solutions in type IIA and
eleven-dimensional supergravities which are expected to be dual to
(2+1)-dimensional Lifshitz-like theories. Under reasonable
ans\"atze, we prove that such gravity duals in the supergravities
are not possible. We also discuss a possible physical reason behind this.

\end{abstract}

\end{titlepage}

\newpage

\section{Introduction}
Recently, two interesting extensions of the AdS/CFT correspondence
\cite{Maldacena} to non-relativistic systems were proposed in the
papers \cite{NRC} and \cite{Ka}. The former has the non-relativistic
conformal symmetry including the special conformal transformation,
while the latter, called quantum Lifshitz-like points, only has the
non-relativistic scale invariance.\footnote{For recent discussions
of the Lifshitz-like points, refer to
\cite{Lifa,Lifb,Lifc,Lifd,Life,Liff,Lifg,Lifh,Lifhh,Lifhhh,ALT,Lifi,Lifj,Lifk,Lifl,Lifm,Lifn,Lifo,Lifp}.
See also the seminal paper \cite{AFF} for the quantum field
theoretic properties of the Lifshitz-like points.}

To understand their microscopic holographic dual theories, we need
to find their concrete string theory realizations. Another reason
for this is that string embeddings are expected to give a
consistency condition for solutions of an effective gravity theory
which lives in the lower dimensions via compactifications
\cite{Swa}. The original gravity duals for the non-relativistic
systems have been proposed in effective gravity-matter theories
living in $d<10$ (or $d<11$) dimension \cite{NRC,Ka}, which are not
a priori guaranteed to be derived from ten (or eleven) dimensional
supergravity. The effective $d$-dimensional spacetime dual to a
$(d-1)$-dimensional Lifshitz-like theory will be called $Li_d$ in
this paper. Later, it has been found that the gravity duals of the
former (i.e. non-relativistic conformal) theories can be realized as
solutions in ten or eleven dimensional supergravities as was first
done in \cite{NRD}. However, the presence of a null circle in the
gravity dual \cite{NRC} of non-relativistic conformal systems makes
the connection to interesting condensed matter systems (such as cold
atoms) ambiguous.

The main purpose of this paper is to point out that it is
surprisingly difficult to embed the gravity duals of quantum
Lifshitz-like points \cite{Ka} into ten or eleven dimensional
supergravities as opposed to our naive expectation. Actually, the
only known supergravity solution with the Lifshitz-like property is
the one in the paper \cite{ALT}. It is based on the D3-D7 system
first introduced in \cite{FLRT} and can be embedded into type IIB
string theory. However, in this solution we can introduce the
anisotropy of the scale transformation only through one of the three
spatial directions therefore it corresponds not to a
non-relativistic quantum Lifshitz point but to a classical Lifshitz
point.\footnote{In the presence of fundamental string sources, we
can also find a type IIB supergravity solution whose metric in the
Einstein frame coincides with $Li_5$ \cite{ANT}. However, in this
solution the dilaton is not constant and it breaks anisotropic scale
invariance again.} Also since it has a non-constant dilaton, the
anisotropic scale invariance only holds at the leading order of
interactions, though thermodynamical quantities such as the entropy
still respect the scale invariance.

In this paper, we will present no-go arguments under certain
ans\"atze of fluxes, which we expect will not cause the loss of
generality significantly. Such ans\"atze are necessary to determine
clearly if physical solutions are possible or not in our analysis.
It may be possible that we can extend our no-go arguments if we develop
excellent technical devises in near future.
Under the ans\"atze
we will prove that no
solutions of the form $Li_4\times M_7$, where $M_7$ is an arbitrary
compact manifold, are allowed in the eleven-dimensional
supergravity even if we take the possibility of warp factor into
account. We will also present a similar no-go argument in the massive
type IIA supergravity. Interestingly enough, we often
encounter unphysical supergravity solutions, whose fluxes become
imaginary-valued.  We will also offer a heuristic physical reason
why quantum Lifshitz-like theories are difficult to realize in
string theory via a holography argument in the final section, which
relates the dual theory to certain non-commutative theories.

Though in this paper we always consider the $(2+1)$-dimensional
Lifshitz-like fixed points to perform concrete calculations, we
believe that the main conclusions can be applied to other
dimensions. As opposed to the proofs of no-go theorems for de-Sitter
spaces \cite{MN} and flux backgrounds \cite{GKP} in supergravity, we
need more detailed arguments due to the lack of the Lorentz
symmetry.

This paper is organized as follows: In section 2, we will review the
gravity dual of quantum Lifshitz-like points following the paper
\cite{Ka}. In section 3, we will prove a no-go theorem for the
spacetime $Li_4\times M_7$ in M-theory assuming a certain ansatz of fluxes.
In section 4, we will present a no-go argument for the
spacetime $Li_4\times M_6$ in massive IIA supergravity under an
even form ansatz of fluxes.
We will also show the no-go theorem for deformations of
standard flux compactifications when $M_6$ is an arbitrary nearly K\"{a}hler
manifold. Moreover, we also consider the possibility of adding an
orientifold 6-plane (O6-plane) and show that the result will not be
changed. In section 5, we will summarize our conclusion and discuss
our results.

\section{Review of Gravity Duals of Lifshitz-like Fixed Points}
\label{sec:dual}  Let us first review the gravity dual $Li_4$ of
Lifshitz-like fixed points in $2+1$ dimension, which was first
presented in \cite{Ka}. We start with the four-dimensional gravity
coupled to a one-form gauge field $A_1$ and a two-form field $B_2$.
The action is \ba S=\int d^4x \s{-g}(R-2\Lambda)-\f{1}{2}\int
\left(F_2\we
*F_2+H_3\we
*H_3\right) -c\int B_2\we F_2. \label{gravil} \ea The gravity dual $Li_4$ of a
Lifshitz-like point is defined by the following metric \ba
ds^2&=&L^2\left(-r^{2z}dt^2+r^2(dx^2+dy^2)+\f{dr^2}{r^2}\right)\\
&\equiv & -(\theta^t)^2+(\theta^x)^2+(\theta^y)^2+(\theta^r)^2, \label{lifour} \ea
where we defined the orthonormal basis of one-forms \be
\theta^t=Lr^z dt,\ \ \ \ \theta^{x}=Lrdx,\ \ \ \ \theta^{y}=Lrdy,\ \
\ \ \theta^{r}=L\f{dr}{r}. \label{norone} \ee The Ricci tensor
$R_{\mu\nu}$ and the Ricci scalar $R_4$ is given by \ba &&
R_{tt}=z(z+2)r^{2z},\ \ \ R_{xx}=R_{yy}=-(2+z)r^2,\ \ \
R_{rr}=-\f{2+z^2}{r^2}. \label{ric}\\
&& R_4=-\f{2}{L^2}(3+2z+z^2).
\ea

In order to break the $(2+1)$-dimensional Lorentz symmetry, we
introduce the background fluxes \ba F_2=\ap \theta^t\we \theta^r,\ \
\ H_3=\beta \theta^x\we \theta^y\we \theta^r. \label{flux} \ea Here
the fact that the fluxes are written by the orthonormal one-forms
(\ref{norone}) guarantees the non-relativistic scale invariance and
is consistent with the Einstein equations discussed below.

In this system the equations of motion for fluxes are given by \be
d*F_2=-cH_3,\ \ \ d*H_3=cF_2. \ee This leads to the constraints \be
\f{\ap}{\beta}=\s{\f{z}{2}}, \ \ \ \ \ \  c^2=\f{2z}{L^2}. \ee The
Einstein equation becomes \be
R_{\mu\nu}+g_{\mu\nu}\left(-\f{R}{2}+\f{1}{24}H_{\rho\sigma\lambda}H^{\rho\sigma\lambda}
+\f{1}{8}F_{\rho\sigma}F^{\rho\sigma}+\Lambda\right)=\f{1}{2}F_{\mu\ap}F_{\nu}^{\,\,\,\ap}
+\f{1}{4}H_{\mu\ap\beta}H_\nu^{\,\,\,\ap\beta}.\label{einh} \ee

By substituting the Ricci tensor (\ref{ric}) and the flux ansatz (\ref{flux}) into (\ref{einh}),
we obtain  in the end
\be
\ap=\f{\s{2z(z-1)}}{L},\ \ \ \ \beta=\f{2}{L}\s{z-1}, \ \ \ \ \Lambda=-\f{z^2+z+4}{2L^2}.
\ee
Notice that in this solution
the coefficient $c$ of Chern-Simons term is uniquely determined by the cosmological constant
$\Lambda$. In the later arguments of its string theory embedding, what turns out to be
 crucial is whether this constraint
is consistent with string theory compactifications or not.

It is also useful to take a duality transformation. We add a term
$\int d\phi\we H_3$ assuming that $H_3$ is now a general three-form.
The equation of motion for $\phi$ leads to the Bianchi identity
$dH_3=0$. If we integrate out $H_3$, we find the action looks like
\ba S=\int d^4x \s{-g}(R-2\Lambda)-\f{1}{2}\int \left(F_2\we
*F_2+(d\phi-cA_1)\we *(d\phi-cA_1) \right). \label{higgs} \ea By absorbing $\phi$
via the gauge transformation, this theory is equivalent to a massive
vector field theory coupled to gravity, as already noted in
\cite{Lifb}.

\section{No-Go Theorem for $Li_4\times M_7$ in M-theory}

\subsection{Direct Product Metric $Li_4\times M_7$}

The action of the eleven-dimensional supergravity is given
by\footnote{ In this paper, we will adopt the normalization of
supergravity in \cite{Pol}.} \be S=\f{1}{2\kappa_{11}^2}\left[\int
d^{11}x\s{-g} \left(R-\f{1}{2\cdot
4!}F_{\mu\nu\rho\sigma}F^{\mu\nu\rho\sigma}\right) -\f{1}{6}\int
C_3\we F_4\we F_4\right]. \ee

The Bianchi identity and equations of motion of the four-form flux
read \be dF_4=0, \ \ \ \ \ d*F_4+\f{1}{2}F_4\we F_4=0, \ee where
$F_4=dC_3$.

We assume that the spacetime metric takes the form of the direct
product $Li_4\times M_7$ without any warp factor, where $M_7$ is
taken to be an arbitrary seven-dimensional compact manifold. The
coordinates of $M_7$ are denoted by $x^i=(x^1,x^2,\ddd,x^7)$. We
denote the volume-form of $M_7$ by $V_7$. Now we take the
the four-form flux with the symmetry of $Li_4$ to be in the following
form
\begin{equation}
F_4=f~ \theta^t \theta^x \theta^y \theta^r+\theta^x\theta^y\theta^r
\we \ap +\theta^t \theta^r \we \beta +\theta^r\we \Omega +\eta.\label{anf}
\end{equation}
This form of flux is the most general form in which we can tell
precisely if the corresponding supergravity solution exists or not,
using the arguments presented below.
Its Hodge dual is given by\footnote{In our convention, the
epsilon tensor is chosen to be
 $\ep_{txyr1234567}=-\ep^{txyr1234567}=1$, which leads to the relation
$\f{1}{p!}F^{\mu_1,\ddd,\mu_p}F_{\mu_1,\ddd,\mu_p}V_7=F\we *F$.}
 \be
*F_4=-f V_7+\theta^t \we *\ap -\theta^x\theta^y\we *\beta +
 \theta^t \theta^x \theta^y \we *\Omega
+\theta^t \theta^x \theta^y \theta^r \we *\eta. \ee  The Einstein
equation requires that the $r$-dependence in $F_4$ only appears
through $\theta^\mu$ as in the previous section. Thus, $f$, $\ap$,
$\beta$, $\Omega$ and $\eta$ are 0,1,2,3 and 4-form in $M_7$ and
they do not depend on $r$.

The Bianchi identity requires \be d\ap=d\beta=d\Omega=df=d\eta=0.
\ee The flux equations of motion lead to \ba
&& d*\ap=d*\beta=d*\Omega=0,\\
&& \beta\we \eta=\f{z}{L}*\ap, \label{bet}\\ && \ap\we
\eta=\f{2}{L}*\beta,   \label{alp}\\
&& d*\eta-\f{z+2}{L}*\Omega+\eta f=0, \\
&& \Omega\we \eta=0. \ea

In particular, from (\ref{bet}) and (\ref{alp}), we find \be
z\ap^2=\beta^2, \label{zab} \ee where we have defined
$\ap^2=\ap_i\ap^i$ and $\beta^2=\beta_{ijk}\beta^{ijk}$ with
$i=1,2,\ddd,7$.

The Einstein equation can be rewritten into the following form \be
R_{\mu\nu}=-\f{1}{6\cdot 4!}F^2g_{\mu\nu}+\f{1}{12}
F_{\mu\rho\sigma\tau}F_{\nu}^{\,\,\,\rho\sigma\tau}.
\label{einno}\ee

Therefore, the Einstein equation (\ref{einno}) for the $(tt)$,
$(xx)$ and $(rr)$ components reads \ba &&
-\f{z(z+2)}{L^2}=-\f{1}{6}(-f^2+\ap^2-\f{1}{2}\beta^2+\f{1}{3!}\Omega^2+\f{1}{4!}\eta^2)
+\f{1}{2}(-f^2-\f{1}{2}\beta^2),\no &&
-\f{2+z}{L^2}=-\f{1}{6}(-f^2+\ap^2-\f{1}{2}\beta^2+\f{1}{3!}\Omega^2
+\f{1}{4!}\eta^2)+\f{1}{2}(-f^2+\ap^2),\no &&
-\f{2+z^2}{L^2}=-\f{1}{6}(-f^2+\ap^2-\f{1}{2}\beta^2+\f{1}{3!}\Omega^2
+\f{1}{4!}\eta^2)+\f{1}{2}
(-f^2+\ap^2-\f{1}{2}\beta^2+\f{1}{6}\Omega^2).\nonumber \ea

By taking differences among them, we can easily find \ba
\f{2(z-1)}{L^2}=\f{1}{2}\ap^2 +\f{1}{12}\Omega^2,\ \ \ \
 \f{z(z-1)}{L^2}=\f{\beta^2}{4}-\f{1}{12}\Omega^2. \ea

By combining these two equations with the relation (\ref{zab}), we
find if $\Omega^2\neq 0$, then $z=-2$. In this case, we only have
unphysical solutions that has $\f{\beta^2}{\ap^2}=-2<0$. Therefore,
we need to require $\Omega^2=0$ (or equivalently
 $\Omega_{ijk}=0$) below.

Now let us look at the off-diagonal components of the Einstein
equation. In particular, if we consider $(ti)$ component it behaves
like $R_{ti}\propto f\ap_i$ and this contradicts our metric ansatz
unless it is vanishing. If we assume $f\neq 0$,  then we need to set
$\ap_i=0$ and $z=1$. Thus we recover
the full Lorentz symmetry, which we do not want.

Thus, we need to set $f=0$. In this case, all equations of
motion can be reduced to \ba &&
\f{2z^2+12z+4}{L^2}=\f{\eta^2}{4!},\label{etas}\\ &&
R_{ij}=-\f{3z}{L^2}g_{ij}
+\f{1}{2}(\ap_i\ap_j-\beta_{ik}\beta_j^k+\f{1}{6}\eta_{iklm}\eta_j^{klm}),\\
&& d\ap=d\beta=d\eta=0,\ \ \ \ \ f=\Omega=0,\ \ \ \ \ d*\ap=d*\beta=d*\eta=0,\\
&& \beta\we \eta=\f{z}{L}*\ap,\ \ \ \ \ \ap\we \eta=\f{2}{L}*\beta,\label{aps}\\
&& \ap^2=\f{\beta^2}{z}=\f{4(z-1)}{L^2}.\label{apss} \ea
We can also find that the value of the Ricci scalar of $M_7$ as
$R_{M}=\f{(2z+3)(z+2)}{L^2}>0$.

By using (\ref{aps}) we can find the relation
\ba \ap\we *\ap =\f{L^2}{2z}~
*(\ap\we\eta)\we
(\ap\we\eta).\label{apa}
\ea
This equation (\ref{apa}) can be rewritten in the component expression as \ba \ap^2&=&\f{25
L^2}{2z\cdot 5!}\ap_{[\mu_1}\eta_{\mu_2\mu_3\mu_4\mu_5]}
\ap^{[\mu_1}\eta^{\mu_2\mu_3\mu_4\mu_5]}\\
&=&\f{L^2}{2z\cdot
4!}\left(\ap^2\eta^2-4\ap^i\ap_j\eta_{iklm}\eta^{jklm}\right). \ea
By plugging into (\ref{apss}) we obtain \be
\f{1}{4!}\ap^i\ap_j\eta_{iklm}\eta^{jklm}=\f{2(z-1)(z^2+5z+2)}{L^4}>0.
\label{etap} \ee On the other hand, we can show $\ap^i\beta_{ij}=0$
as we can easily see $\ap\we *\beta=0$ from (\ref{aps}). By
combining these results, we can finally evaluate \be
R_{ij}\ap^i\ap^j=\f{4z(z-1)(z+4)}{L^4}>0, \label{rij} \ee which
actually proves that there is no such harmonic one-form on $M_7$.
This statement follows from the Weitzenbock formula (see e.g.
\cite{Einstein}) \be \ap^i \Delta
\ap_i=-\ap^i\nabla_j\nabla^j\ap_i+R_{ij}\ap^i\ap^j,\label{wei} \ee
where the Laplacian is defined by $\Delta=d\delta+\delta d$ as usual
in the harmonic analysis; $\delta$ is the codifferential and
$\nabla$ is the covariant derivative. Since a harmonic one-form
$\ap$ satisfies $d\ap=\delta \ap=0$, the left-hand side of
(\ref{wei}) vanishes. On the other hand, if we integrate the
right-hand side on $M_7$, it should be positive by the partial
integration and the formula (\ref{rij}). Therefore, such a harmonic
one-form $\ap$ cannot exist and this completes our no-go argument
for $Li_4\times M_7$.

\subsection{Warp Factor in $Li_4\times M_7$}

Actually, we can still have the freedom of taking the warp factor of
the metric into account as follows  \be
ds^2=e^{2A(\xi)}ds^2_{Li_4}+e^{\f{4}{5}A(\xi)}ds^2_{M_7}, \ee where
$\xi$ denotes the seven coordinates of $M_7$. We fixed the power of
the warp factor by the reparameterizations such that the one-form
$\ap$ in $M_7$ again becomes harmonic $d\ap=d*\ap=0$ due to the flux
equation of motion. The equations of motion for fluxes are only
modified by powers of $e^A$. For example, we can show \be
\ap^2=\f{\beta^2}{z}e^{\f{6}{5}A}=4(z-1)e^{\f{24}{5}A}. \ee We also
have to require $f=\Omega_{ijk}=0$ again. Moreover, by using these
relations, we can eventually evaluate the right-hand side of
(\ref{wei}) as follows: \ba &&
R_{ij}\ap^i\ap^j-\ap^i\nabla_j\nabla^j\ap_i\no
&=&4z(z-1)(z+4)e^{\f{18}{5}A}+\f{48(z-1)}{5}\left(\nabla_i\nabla^i
A+6(\nabla_i A)(\nabla^i
A)\right)e^{\f{24}{5}A}+6\ap^i\ap^j\nabla_i\nabla_j A\no
&&+(\nabla_i\ap^j)(\nabla^i\ap_j)-\nabla_j(\ap^i\nabla^j\ap_i)\no
&=&4z(z-1)(z+4)e^{\f{18}{5}A}-\f{1152}{25}(z-1)(\nabla_i A)(\nabla^i
A)e^{\f{24}{5}A}
+6\nabla^i(\ap_i\ap_j\nabla^jA)+(\nabla_i\ap^j)(\nabla^i\ap_j).\nonumber\\
\label{erqi} \ea

Now let us recall that $M_7$ is a compact manifold. Therefore, the
function $A(\xi)$ on $M_7$ should take its maximum and minimum value
somewhere. Suppose it takes the minimum value at
 $\xi=\xi_{min}$. It is obvious that
\be \de_iA(\xi_{min})=0,\ \ \ \ \ \
\ap^i\ap^j\de_i\de_jA(\xi_{min})\geq 0. \ee On the other hand, since
$\ap$ is a harmonic one-form, the left-hand side of (\ref{erqi}) is
vanishing as explained before. By evaluating this at
$\xi=\xi_{min}$, we obtain \be
0=4z(z-1)(z+4)e^{\f{18}{5}A}+6\ap^i\ap^j\de_i\de_j
A+(\nabla_i\ap^j)(\nabla^i\ap_j). \ee This is clearly inconsistent
as its right-hand side is positive. Thus, there are no such harmonic
one-form $\ap$ and this finishes the no-go theorem for the warped
case.

\section{Massive IIA Supergravity on $Li_4\times M_6$}

\subsection{Massive IIA theory}

The massive IIA supergravity \cite{Romans} is obtained by adding a
new auxiliary scalar field $M$ and ten-form RR-flux $F_{10}=dC_9$
and by shifting the flux $F_2$ and $F_4$ such that \ba &&
\ti{F}_2=dC_1-MB_2,\no && \ti{F}_4=dC_3-A_1\we H_3+\f{1}{2}MB_2\we
B_2. \ea The action is defined by \be S_{massive\
IIA}=\f{1}{2\kappa^2}\int {\cal L}, \ee where \ba && {\cal
L}=\s{-g}e^{-2\phi}(R+4\de_\mu\phi\de^\mu\phi)-\f{e^{-2\phi}}{2}H_3\we
*H_3 -\f{1}{2}\ti{F}_2\we *\ti{F}_2-\f{1}{2}\ti{F}_4\we
*\ti{F}_4-\f{1}{2}M\we *M \no && \ +M\we F_{10}-\f{1}{2}B_2\we
dC_3\we dC_3-\f{M}{6}dC_3\we B_2\we B_2\we B_2-\f{M^2}{40} B_2\we
B_2\we B_2\we B_2\we B_2.\nonumber \ea

The Bianchi identities and equations of motion of the fluxes are
summarized as follows: \ba && dH_3=0,\ \ \ \ dM=0,\no &&
d\ti{F}_2=-MH_3,\ \ \ d\ti{F}_4=-\tilde{F}_2\we H_3,\no &&
d*\ti{F}_2=H_3\we
*\ti{F}_4,\no && d*\ti{F}_4=-\ti{F}_4\we H_3,\no &&
d(e^{-2\phi}*H_3)=-M\we
*\ti{F}_2+\f{1}{2}\ti{F}_4\we \ti{F}_4-\ti{F}_2\we *\ti{F}_4. \ea

The equation of motion for $C_9$ requires that $M$ is a constant:
$M=m$. On the other hand, the equation of motion of $M$ just
determines $F_{10}$ in terms of other fields.

The dilaton equation of motion is
\be
R+4\nabla_\mu\nabla^\mu \phi-\f{1}{12}H_{\mu\nu\rho}H^{\mu\nu\rho}
-4\nabla_\mu\phi\nabla^\mu\phi=0.
\label{dilat}
\ee

The Einstein equation becomes \ba
R_{\mu\nu}+\f{1}{4}g_{\mu\nu}A=-2\nabla_\mu\nabla_\nu\phi
+\f{1}{4}H_{\mu\ap\beta}H_{\nu}^{\,\,\,\ap\beta}+\f{1}{2}
e^{2\phi}\ti{F}_{\mu\ap}\ti{F}_\nu^{\,\,\,\ap}+\f{1}{12}e^{2\phi}
\ti{F}_{\mu\ap\beta\gamma}\ti{F}_\nu^{\,\,\,\ap\beta\gamma},\label{Eins}
\ea where \ba
A\equiv\f{e^{2\phi}}{4!}\ti{F}_{\mu\nu\rho\sigma}\ti{F}^{\mu\nu\rho\sigma}
+\f{e^{2\phi}}{2!}\ti{F}_{\mu\nu}\ti{F}^{\mu\nu}+e^{2\phi}M^2. \ea

\subsection{No-go Argument in Massive IIA on $Li_4\times M_6$}

Now we would like to see if the spacetime $Li_4\times M_6$ can be a
classical solution of the ten-dimensional massive IIA supergravity.
The form of Ricci tensor and the scale invariance require that the
dependence of fluxes on $r$ all comes from the normalized one-forms
$\theta^{t,x,y,z}$ again.

In this subsection, we take into account only even forms in $M_6$ in
the flux ansatz to make the problem easier. In this case, the general
flux ansatz with the required Lorentz symmetry breaking for
the Lifshitz spacetime can be written as follows\footnote{We can show
from equations of motion that the term like $H_3\sim \theta^r\we J$ is
not allowed.}
\ba
&& M=m,\no
&& \ti{F}_2=\ap\theta^t \theta^r+\eta\theta^x\theta^y+J_1,\no
&& H_3=\beta\theta^x  \theta^y \theta^r,\no
&& \ti{F_4}=f\theta^t\theta^x\theta^y\theta^r+ \theta^x\theta^y
J_2+\theta^t\theta^r J_3+V_4, \label{anfl} \ea where $\ap,\beta$ and $\eta$ are
constants. $J_1,J_2$ and $J_3$ are some two-forms on $M_6$, and
$V_4$ is a certain four-form on $M_6$.

Bianchi identities lead to \ba dJ_1=dJ_2=dJ_3=0,\ \ \ \ \
J_2=-\f{\beta L}{2}J_1, \ \ \ \ \ m=-\f{2\eta}{\beta L}, \ \ \ \ \
dV_4=0. \ea The flux equations of motion for $\ti{F}_2$ and
$\ti{F}_4$ lead to \ba d*J_1=d*J_3=0, \ \ \ \ d*V_4=0,\ \ \ \ f\beta
L=2\ap,\ \ \ \ \beta L V_4=2*J_3. \ea Below we write $J_1\equiv J$
and $J_3\equiv K$. Notice that $J_2=-\f{\beta L}{2}J$ and
$V_4=\f{2}{\beta L}*K$.

Notice that the presence of a non-trivial warp factor $\Omega(\xi)$
(which depends on the coordinates $\xi$ of $M_6$) in the metric
$ds^2=\Omega^2(ds^2_{Li_4}+ ds^2_{M_6})$ contradicts with one of the
flux equations of motion that reads $f\beta L=2\ap\Omega^4(\xi)$.
Therefore, in this setup we can set $\Omega=1$, namely the metric is
in the form of the direct product.

If we plug in the explicit values of the fluxes, the $(tt)$, $(xx)$
and $(rr)$ components of Einstein equation are expressed as follows:
\ba
&&-\f{z(z+2)}{L^2}=-\f{A}{4}-\f{g_s^2}{2}\ap^2-\f{2g_s^2\ap^2}{\beta^2L^2}
-\f{g_s^2K^2}{4}, \label{einz}\\
&&-\f{z+2}{L^2}=-\f{A}{4}-\f{2g_s^2\ap^2}{\beta^2L^2}+\f{g_s^2\beta^2 L^2 J^2}{16}
+\f{g_s^2 m^2L^2\beta^2}{8}
+\f{\beta^2}{2},\\
&&-\f{z^2+2}{L^2}=-\f{A}{4}-\f{g_s^2}{2}\ap^2-\f{2g_s^2\ap^2}{\beta^2L^2}
-\f{g_s^2K^2}{4}+\f{\beta^2}{2},
\ea
with
\be
A=g_s^2\left[-\f{4\ap^2}{\beta^2L^2}+\f{J^2\beta^2L^2}{8}-\f{K^2}{2}
+\f{2K^2}{\beta^2L^2}-\ap^2
+\f{m^2L^2\beta^2}{4}+\f{J^2}{2}+m^2\right].
\ee

By taking the differences, we immediately obtain
\ba
&& \beta^2=\f{4(z-1)}{L^2},\label{eemo} \\
&& \f{z(z-1)}{L^2}=\f{g_s^2m^2L^2\beta^2}{8}+\f{g_s^2\ap^2}{2}
+\f{g_s^2\beta^2L^2J^2}{16}+\f{g_s^2 K^2}{4} \label{eemt}. \ea

On the other hand, after combined with (\ref{eemo}) and
(\ref{eemt}), the equation (\ref{einz}) actually leads to  $z=-4$.
This shows that we can only have unphysical solutions as the
equation (\ref{eemo}) implies that the $H$-flux becomes imaginary.
In this way, we have found that under this ansatz, there is no
physical solution corresponding to the Lifshitz-like fixed point in
the IIA massive supergravity.

Moreover, we can find explicit unphysical solutions. Suppose that
$M_6$ is K\"{a}hler-Einstein and that the fluxes are given by the
parameters \ba m=\eta=0,\ \ \ J_1=-\f{2k}{R^2}\omega, \ \ \ \
J_2=\f{kL\beta}{R^2}\omega,\ \ \ J_3=V_4=0, \ea where $R$ is the
radius of $M_6$ such that $R_{ij}=\f{8}{R^2}g_{ij}$. The K\"ahler
form $\omega$ is a harmonic two-form, which satisfies
$g^{kl}\omega_{ik}\omega_{jl}=g_{ij}$. Then we can show that all
equations of motion are satisfied if we set \be
\ap^2=\f{35}{2g_s^2L^2},\ \ \ \beta^2=-\f{20}{L^2},\ \ \ \
k^2=-\f{6L^2}{g_s^2},\ \ \ R^2=4L^2,\ \ \ z=-4. \ee This requires
that $\beta$ and $k$ to be imaginary. This argument includes
Lorentz symmetry breaking deformations of the
$AdS_4\times CP^3$ solution dual to the ${\cal N}=6$ Chern-Simons gauge theory
 \cite{ABJM}. Notice that this is a rather basic setup as in this model all moduli are
stabilized.

Finally, it is also possible to check that the situation will not be
improved even if we insert spacetime filling D-branes i.e. non-BPS
D9-branes\footnote{For the definition of non-BPS D-branes refere to
e.g. the review article \cite{Sen}.} whose action is proportional to
$\int d^9x e^{-\phi}\s{g}$ setting the tachyon field vanishing.

\subsection{No-go Argument for Nearly K\"{a}hler Compactifications}

To generalize our previous no-go argument of Lifshitz-like
solutions, we need to take into account odd forms in the compact
six-dimensional manifold $M_6$. Since a general argument in this case
turns out to be quite complicated, we would like to assume the case
where $M_6$ is a nearly K\"{a}hler manifold, including Calabi-Yau
manifolds. This is motivated by the expectation that the
Lifshitz-like solutions will appear by deforming the $AdS_4\times
M_6$ solution to the massive IIA supergravity. The backgrounds
$AdS_4\times M_6$ have been intensively studied in the context of
flux compactifications (see e.g. \cite{LuTs,DGKW,ABV,AF,KaP}). One
major way to maintain the $N=1$ supersymmetry after the
compactification is to assume that $M_6$ is a nearly K\"{a}hler
manifold \cite{LuTs,AF,KaP}, which is known to be Einstein. Thus, we
believe that this restriction to nearly K\"{a}hler manifolds will
not lose the generality seriously.

When $M_6$ is nearly K\"{a}hler, the two-form $J$ that defines the
almost complex structure and the holomorphic three-form $\Omega$
satisfies\ba
&&\Omega \we J=0,\\
&&V_6=\text{Re}\Omega\we \text{Im}\Omega=\f{1}{3!}J\we J\we J,\\
&&d \text{Re} \Omega=0,\\
&&dJ=b \text{Re} \Omega,\\
&&d \text{Im}\Omega=-\f{b}{6}J\we J. \ea The constant $b$ measures
the extent to which the metric deviates from the Calabi-Yau
condition and is relate to the Ricci curvature via \be
R_{ij}=\f{5}{36}b^2 g_{ij}. \label{rbr} \ee Since it is
straightforward to show that there is no physical solution when
$b=0$ (i.e. when $M_6$ is Calabi-Yau), we will restrict to the
values $b\neq 0$ below.

By employing $J$ and $\Omega$, we can write down the flux ansatz\footnote{
In this convention the known $AdS_4\times M_6$ solution \cite{LuTs,AF,KaP} corresponds to
$z=1,\ \  \eta=g=\ap=\beta=h=q=0,\ \  a=-\f{m}{\s{15}},\ \  s=\f{3}{5}m, \ \
f=-\f{3\s{3}}{\s{5}}m,\ \  b=\f{\s{15}}{L},\ \  k=\f{1}{L}$.} with the
 expected Lorentz symmetry breaking as follows:\footnote{
Here the term $\theta^r\we J$ is not allowed in $H_3$ as we assume
$dJ\neq 0$; the term like
 $\theta^r\we \text{Re}\Omega$ is not included in $\ti{F}_4$ as it is not consistent with
 the equation of motion $d*F_2=H_3\we *\ti{F}_4$.}
\ba
&&M=m,\no
&&\ti{F}_2=aJ+\ap\theta^t\theta^r+\eta\theta^x\theta^y,\no
&&H_3=\beta\theta^x\theta^y\theta^r+k \text{Re}\Omega,\no
&&\ti{F_4}=f\theta^t\theta^x\theta^y\theta^r+g\theta^x\theta^y\we
J+h\theta^t\theta^r\we J +q\theta^t\we \text{Re}\Omega+\f{s}{2}J^2. \label{ansfll} \ea

The Bianchi identities lead to \ba
&&\eta=-\f{m\beta L}{2},\ \ \ ab+km=0, \label{ftn} \\
&& g=-\f{\beta a L}{2},\ \ \ gb+\eta k=0,\ \ \ hb-\f{zq}{L}+\ap k=0.
\label{ffn} \ea

The equations of motion lead to \ba
&& \f{2\ap}{L}=f\beta +kq,\\
&& kf-\beta q+sb=0,\ \ \ \  -\f{h}{L}+\f{bq}{6}+\f{s\beta}{2}=0,\\
&& \f{z\beta}{Lg_s^2}+m\eta-3hs+3ga-\ap f=0,\\
&& \f{ma}{2}-\f{fs}{2}-gh+sa-\f{\ap h}{2}+\f{g\eta}{2}-\f{bk}{6g_s^2}=0,\label{hfe} \\
&& m\ap+3gs+3ha+f\eta=0. \ea

Next we would like to write down the Einstein equations\footnote{
Notice that we do not have to worry the unwanted cross terms in the
Einstein eq. such as $R_{ti}$ and $R_{ri}$ as
$J_{ij}\Omega^{ij}_k=0$.}. The useful identities are \be
J_{ik}J_{jl}g^{kl}=g_{ij},\ \ \ \
\text{Re}\Omega_{ikl}\text{Re}\Omega_j^{\,\,\,kl}=g_{ij}. \ee

The Einstein equation can be eventually expressed as follows after taking suitable linear combinations
\ba
&& -\f{z(z+2)}{L^2}=-\f{g_s^2}{4}(\ap^2+f^2+3g^2+3h^2+q^2+3s^2+3a^2)-
\f{g_s^2\eta^2}{4}-\f{g_s^2m^2}{4},\label{eino}\\
&& \f{z(z-1)}{L^2}=\f{g_s^2}{2}(\ap^2+3h^2+3g^2)+\f{g_s^2\eta^2}{2},\\
&& \f{2(z-1)}{L^2}=\f{g_s^2}{2}q^2+\f{\beta^2}{2}.\label{qbe} \ea

If $q=0$, then the situation becomes simpler and become the same as
the previous subsection and we immediately find $z=-4$. In this
case, there are no physical solutions. Thus, we need to set $q\neq
0$ below.

Now the remaining equations of motion which we have to impose
are the dilaton equation and the Einstein
equation in $M_6$ direction. They are equivalent to
\ba && -\f{2(z^2+2z+3)}{L^2}+R_6=\f{\beta^2+k^2}{2},\\
0&=&\f{1}{6}H_{\mu\nu\rho}H^{\mu\nu\rho}-\f{3g_s^2}{4}
\ti{F}_{\mu\nu}\ti{F}^{\mu\nu}-\f{g_s^2}{48}\ti{F}_{\mu\nu\rho\lambda}
\ti{F}^{\mu\nu\rho\lambda}-\f{5}{2}g_s^2m^2\no
&=&\beta^2+k^2-\f{3g_s^2}{2}(3a^2-\ap^2+\eta^2)-\f{g_s^2}{2}
(-f^2+3g^2-3h^2-q^2+3s^2)-\f{5}{2}g_s^2m^2.\nonumber
\ea

By combining these with previous equations we can actually show \be
3L^2k^2+3L^2\beta^2=-6+z-2z^2. \ee Since the right-hand side is
clearly negative when $z>1$, we cannot construct any physical solutions.

\subsection{Adding Orientifold Plane}

Adding orientifold planes in general enlarges the set of possible
solutions in the flux compactification. For example, adding an O6
plane to the IIA string flux compactification allows supersymmetric
$AdS_4\times M_6$ solutions with all moduli stablized and with $M_6$
being half-flat \cite{LuTs,DGKW,ABV,AF,KaP}. Supersymmetry requires
an O6 to wrap the entire $AdS_4$ and a supersymmetric three-cycle
inside $M_6$. Moreover, smearing an O6 inside $M_6$ allows solutions
with $M_6$ being Calabi-Yau three-fold. For our present purpose,
since we do not require Lorentz symmetry, more generic O6
configurations should in principle be allowed (for example, with an
O6 wrapping only $(txy)$-direction and a four-cycle inside $M_6$).
However, for simplicity and for parallel comparison with the $AdS_4$
case, we will only consider the configuration of an O6 wrapping the
entire $Li_4$ and a three-cycle inside $M_6$.

In this subsection, we will show that adding an O6 plane wrapping
the entire $Li_4$ and a three-cycle inside $M_6$  will not help
evade the no-go theorem in the case of $Li_4 \times M_6$ with $M_6$
being nearly K\"ahler. To warm up, we first show that even in the
presence of a smeared O6 plane, there exist no Lifshitz-like
solutions when $M_6$ is a Calabi-Yau three-fold with only one
two-form (the K\"ahler form). Then, we will generalize the result to
the case where $M_6$ is a nearly K\"ahler manifold.

The O6 plane couples to the $C_7$ flux and sources $F_2$
magnetically.  Therefore, in the presence of O6 plane, the Bianchi
identity of $F_2$ flux becomes
\begin{equation}
d\ti{F}_2=-MH_3-\mu_6 \delta_3
\end{equation}
where $\delta_3$ is three-form localized on the supersymmetric
three-cycle. The smearing of the O6 plane is simply done by
replacing $\mu_6\delta_3$ with $\mu_6' \mbox{Re}\Omega$ \cite{DGKW}.

For $M_6$ being $CY_3$ with only one two-form, we can first start
with the most generic flux ansatz, then use the Bianchi identities
and equations of motions of the fluxes to eliminate most of the
terms. In the end, there are only two possibilities:
\begin{eqnarray}
(1)&& H_3=\beta \theta^{xyr}, \qquad
\tilde{F}_2=\alpha\theta^{tr}+\eta\theta^{xy}+ a J, \qquad
\tilde{F}_4=f \theta^{txyr} +\frac{s}{2}J^2 +h \theta^{tr}\wedge J
+g
\theta^{xy}\wedge J \non\\
(2) &&H_3=k\mbox{Re}\Omega, \qquad \tilde{F}_2=a J, \qquad
\tilde{F}_4=\frac{s}{2} J^2
\end{eqnarray}
The first one reduces to the situation without the O6 plane. Only
the second one takes advantage of the presence of the O6 plane,
however, it does not break Lorentz symmetry. Therefore, adding an O6
plane in this case does not help finding solution with the
anisotropic scale invariance. This is also clear from (\ref{qbe}) as
it leads to $z=1$ by setting $q=\beta=0$. Notice that though the
presence of O6 plane changes Einstein equations, it does not change
the differences of them such as (\ref{qbe}).

For $M_6$ being nearly K\"ahler, the ineffectiveness of adding an O6
also arises from the fluxes. The flux ansatz remains the same as the
one without the O6; the only change is that the second equation
coming from the $F_2$'s Bianchi identity now becomes
\begin{equation}
ab+km=n_6,
\end{equation}
where $n_6$ counts the smeared O6 charge after a proper
normalization. In the process of solving flux Bianchi identities and
equations of motion, the first equation from $F_4$'s Bianchi
identity and the third equation from $H_3$'s equations of motion
become
\begin{equation}
\beta n_6=0, \qquad \textrm{and} \qquad qn_6=0,
\end{equation}
respectively. Namely, for a nonzero smeared O6 charge, we have
$q=\beta =0$. Again (\ref{qbe}) requires the relativistic dynamical
exponent $z=1$. This means that adding an O6 does not help
evade the no-go theorem in the present setup.

\section{Conclusion and Discussion}
\label{sec:conc}

\subsection{Conclusions: No-go Theorems for Lifshitz-like Spacetime in Supergravities}
In this paper, we examined possibilities of string theory embeddings
of the gravity duals (denoted by $Li_d$) for $(d-1)$-dimensional
quantum Lifshitz-like fixed points, which are invariant under
anisotropic scale transformations. We considered the ten- or
eleven-dimensional supergravity description of string theory by
taking the ordinary low energy limit. A no-go argument in
supergravities as general and simple as the one for de-Sitter spaces
\cite{MN} or the one for flux compactifications \cite{GKP} does not
seem to be available in our case because we consider less symmetric
spacetimes with a negative curvature. Therefore, we had to examine
each possibilities of compactifications of supergravities.

In all of our supergravity setups (we will summarize them in the next subsection)
we examined, we found that it
is impossible to construct $Li_4$ solutions.  Notice that in all such
setups, there exist $AdS_4$ solutions. Even though our analysis has
been done only for the $Li_4$ spacetime defined by (\ref{lifour}),
we believe that our result will not significantly lose generality as
many aspects of $AdS_4$ solutions are similar to $AdS_3$ and
$AdS_5$. Motivated by these results, we are tempted to conjecture
that the gravity duals $Li_d$ of Lifshitz-like theories cannot be
embedded into any supergravity description of string theory or
M-theory. The rigorous proof of this no-go theorem will certainly be
an important future problem.

Under certain flux ansatze,
we proved that no solutions of the form $Li_4\times
M_7$, where $M_7$ is an arbitrary compact manifold, are allowed in
the eleven-dimensional supergravity, even if we take the possibility
of warp factor into account. We also presented a similar no-go
theorem in the massive type IIA supergravity for the spacetime
$Li_4\times M_6$. Finally we examined massive
IIA compactifications on nearly K\"{a}hler manifolds and showed that
solutions of the form $Li_4\times M_6$ are impossible.

Since the Einstein gravity coupled to a massive vector field can
have a Lifshitz-like solution as we noticed in the end of section 2,
one may expect that holographic superconductor systems
\cite{SCA,SCB} which can be embedded into string theory may have
$Li_d$ solutions.\footnote{We would like to thank Gary Horowitz very
much for useful discussions on this possibility.} However, we can
check after some analysis that $Li_d$ solutions are impossible as
least in the recently found string theory embeddings \cite{SCC,SCD}
via ten or eleven dimensional supergravity, as consistent with our
claim. In the Chern-Simons-Maxwell type description (\ref{gravil})
of the gravity dual, our no-go theorem comes from the fact that in
the string compactification, the cosmological constant $\Lambda$ and
the Chern-Simons coefficient $c$ cannot be chosen independently.

As in the case of de-Sitter spaces \cite{KKLT} or flux
compactifications \cite{GKP}, one might expect that the addition of
orientifolds or non-supersymmetric D-branes may evade our no-go
arguments in supergravities. We performed preliminary analysis in
tractable examples and observed that such effects do not help us to
construct $Li_d$ spacetimes. Therefore, we do not know at all
whether a gravity dual of non-relativistic Lifshitz-like theories
exists in string theory or not. We might also improve this situation
by considering non-critical string setups such as in \cite{ASS}.
These issues should certainly deserve future studies.

\subsection{Summary of the No-go Setups}

Since it might be helpful for future studies of possible Lifshitz-like solutions in 
supergravities, we would like to summarize our setups where we could manage to prove the no-go 
theorems for a four dimensional Lifshitz spacetime $Li_4$ as follows:
\begin{itemize}
  \item The eleven-dimensional supergravity with the four form flux ansatz
given by (\ref{anf}) compactified on an arbitrary manifold $M_7$. (Section 3)

  \item  The ten-dimensional massive IIA supergravity with the flux ansatz 
  (\ref{anfl}) compactified on an arbitrary manifold $M_6$. (Section 4.2)
  
  \item  The ten-dimensional massive IIA supergravity with the flux ansatz
  (\ref{ansfll}) compactified on an nearly Kahler manifold $M_6$ . (Section 4.3)
  
  \item The previous massive IIA nearly Kahler compactifications with orientfold 6-plane.
  (Section 4.4)
  
\end{itemize}

\subsection{Why String Duals of Lifshitz-like Points are Hard}

Before we end this paper, we would like to suggest an intuitive
physical reason for our no-go theorems. Consider type IIA or IIB
string theory. Let us remember that to realize a $Li_4$ solution we
need to turn on several RR and $H$ fluxes to break the Lorentz
symmetry. The presence of the Chern-Simons coupling, which is
important in the argument of \cite{Ka} as reviewed in section 2,
requires turning on an $H$ flux. Then the requirement $dH=0$ argues
that $H\propto \theta^x\theta^y\theta^r$ as is indeed so in our
setups.\footnote{ Notice that the solution \cite{ALT} has vanishing
$H$-flux. This is the reason why we managed to find the physical
solution. However, in this solution the dilaton depends on $r$ and
the metric becomes identical to $Li_5$ only in the Einstein frame.
Thus it is out of the ansatz in our paper since we require the
rigorous anisotropic scale invariance.}
 We can easily imagine that such a gravity solution should
be dual to a non-commutative gauge theory as the NS $B$ field will
be induced in the boundary \cite{SW}. The effect of
non-commutativity is interpreted as the the UV cutoff of the gauge
theory. Therefore, the AdS/CFT argues that the IR region is
described by $AdS$,
 while in the UV region, the spatial directions shrink to
zero size. The latter fact requires $z<0$ as in the solutions
 \cite{NCYM} and we will not have a nice scaling limit which extracts the UV part because
it contradicts with the relation like (\ref{apss}), which requires $z>1$.
 Note that a simple analysis shows that this
condition $z>1$ should always be correct even in the presence of the warp
factor in the metric $Li_4\times M_6$.

\vskip3mm

\noindent {\bf Acknowledgments}

We are very grateful to T. Azeyanagi, T. Banks, G. Giribet, J.
Gomis, A. Maloney, J. Mcgreevy, M. Mulligan, N. Ohta,  M. Schulz, 
D. T. Son, A. Strominger and K. Yoshida for useful
discussions, and especially to
G. Horowitz, S. Kachru and Y. Nakayama for very important comments.
 We are supported by World Premier International Research
Center Initiative (WPI Initiative), MEXT, Japan.
 The work of TN is supported by JSPS Grant-in-Aid for
Scientific Research No.19$\cdot$3589.
 The work of TT is supported in part by JSPS Grant-in-Aid for
Scientific Research No.20740132 and by JSPS
Grant-in-Aid for Creative Scientific Research No. 19GS0219.
TT would like to thank very much the organizers and participants
in the workshop ``Quantum Criticality and the AdS/CFT Correspondence'',
at KITP in UCSB and those in the conference ``Quantum Theory and Symmetries 6''
at University of Kentucky, where some parts of our work have been done.

\vskip2mm

%%%%%%%%%%%%%%%%%%%%%%%%%%%%%%%%%%%%%%%%%%%%%%%%%%%%%%%%%%%%%%%%%%%%%%%%%%%%%%%%

%%%%%%%%%%%%%%%%%%%%%%%%%%%%%%%%%%%%%%%%%%%%%%%%%%%%%%%%%%%%%%%%%%%%%%%%%%%%%

\end{document}